\documentclass[12pt]{article}
\usepackage{cite}
\usepackage{amsfonts}
\usepackage{amssymb}
\usepackage{amsthm}
\usepackage{amsmath}
\usepackage{pdfsync}
\usepackage{enumerate}
\usepackage{placeins}
\usepackage{lscape}
\usepackage[pdftex,bookmarks,colorlinks,pdfstartview=FitH]{hyperref}
\hypersetup{colorlinks,%
citecolor=blue,%
filecolor=black,%
linkcolor=blue,%
urlcolor=black,%
pdftex}

\title{\textbf{Conformal Collineations of the Ricci and Energy-Momentum Tensors in Static Plane Symmetric Spacetimes}}
\date{}
\theoremstyle{definition}

\newtheorem{exmp}{Example}[section]
\numberwithin{equation}{section}

\author{ \small  Sumaira Saleem Akhtar$^\text{a}$,\ Tahir Hussain$^\text{a}$\footnote {Corresponding Author Email: tahirhussain@upesh.edu.pk}, \ Ashfaque H. Bokhari$^\text{b}$\ and Fawad Khan$^\text{a}$}
\begin{document}
\maketitle
\begin{scriptsize}
\noindent$^\text{a}$Department of Mathematics, University of Peshawar, Khyber Pakhtoonkhwa, Pakistan.\\
\noindent$^\text{b}$Department of Mathematics and Statistics, King Fahd University of Petroleum and Minerals, Dhahran 31261, Saudi Arabia.
\end{scriptsize} \\
%------------------------------------------------------
\begin{center}
\noindent\textbf{\large Abstract}
\end{center}
%\begin{footnotesize}
Considering the degenerate and non-degenerate cases, we provide a complete classification of static plane symmetric spacetimes according to conformal Ricci collineations (CRCs) and conformal matter collineations (CMCs). In case of non-degenerate Ricci tensor, a general form of vector field generating CRCs is found in terms of unknown functions of $t$ and $x$, subject to some integrability conditions. The integrability conditions are then solved in different cases depending upon the nature of Ricci tensor and it is concluded that static plane symmetric spacetimes possess 7, 10 or 15-dimensional Lie algebra of CRCs. Moreover, it is found that these spacetimes admit infinite number of CRCs when the Ricci tensor is degenerate. A similar procedure is adopted for the study of CMCs in degenerate and non-degenerate matter tensor cases. The exact form of some static plane symmetric spacetimes metrics is obtained admitting non-trivial CRCs and CMCs. Finally, we present some physical implications of our obtained results by considering a perfect fluid as a source of the energy-momentum tensor. \\\\
%\end{footnotesize}
\begin{footnotesize}
\end{footnotesize}
\section{Introduction}
In General Relativity (GR), symmetries are vector fields that preserve certain interesting features of spacetime geometry, some of which include metric, Riemann, stress-energy and Ricci tensors. More specifically, these symmetries when applied to tensors other than metric tensor are generally known as collineations. Mathematically, collineations are vector fields $\xi$ such that the Lie derivative of some arbitrary tensor $\Omega$ vanishes in the direction of $\xi,$ that is \cite{[1]}:
\begin{equation}
\mathcal{L}_\xi \Omega=0  \label{eq:(1.1)}
\end{equation}
where $\mathcal{L}_\xi$ signifies Lie derivative operator along the vector field $\xi.$ If $\Omega$ denotes the metric tensor, then the above equation becomes the well known Killing's equation $\mathcal{L}_\xi g_{ab}=0$ whose solution are called Killing vectors (KVs) admitted by a spacetime metric. The Ricci collineations (RCs) are obtained by taking $\Omega=R_{ab}$ in Eq. (\ref{eq:(1.1)}), where $R_{ab}$ represents the Ricci tensor. One can similarly define curvature (CCs) and matter collineations (MCs) by replacing $\Omega$ in Eq. (\ref{eq:(1.1)}) with Riemann and stress-energy tensors respectively.\\
In addition, if the flow is induced by a vector field $\xi,$ then the metric $g_{ab},$ the Ricci tensor $R_{ab}$ and correspondingly the stress-energy tensor $T_{ab}$ preserves only its conformal class, and hence the $\xi$ is known as a conformal Killing vector (CKV), a CRC or a CMC respectively. Such type of collineations do not necessarily preserve the metric, Ricci or stress-energy tensors. In particular the CRCs satisfy the relation \cite{[2]}:
\begin{equation}
\mathcal{L}_\xi R_{ab}=R_{ab,c}\ \xi^c+R_{bc}\ \xi^c_{,a}+R_{ac}\ \xi^c_{,b}= 2 \lambda R_{ab}, \label{eq:(1.2)}
\end{equation}
where $\lambda$ is a smooth real valued function on the spacetime $M,$ known as conformal factor. If $\lambda$ is a constant, then a CRC is called a homothetic Ricci collineation (HRC) and if $\lambda=0,$ CRCs reduce to RCs. The CMCs are similarly defined by replacing the Ricci tensor $R_{ab}$ in Eq. (\ref{eq:(1.2)}) by the energy-momentum tensor $T_{ab}$. Similar remarks follow for CKVs, which are obtained by replacing $R_{ab}$ with $g_{ab}$ in Eq.(\ref{eq:(1.2)}). If $\xi$ is both a CKV with conformal factor $\alpha$ and a CRC with a conformal factor $\lambda,$ then $\alpha$ and $\lambda$ are related as \cite{[3]}:
\begin{equation}
\alpha_{;ab}=-\lambda \left(R_{ab}-\frac{R}{6} g_{ab}\right).                \label{eq:(1.3)}
\end{equation}
The sets $CRC(M)$ and $CMC(M)$ consisting of all CRCs and CMCs respectively of a spacetime metric representing a manifold $M$ constitute vector spaces which may be of infinite dimension in case of degenerate Ricci or energy-momentum tensor. However, in non-degenerate case, these sets  of collineation vectors form  finite-dimensional Lie algebras such that $dim \ CRC(M) \leq 15$ and $dim \ CMC(M) \leq 15.$ The upper bound of the dimensions of these Lie algebras is attained if and only if the spacetime is conformally flat \cite{[4]}.\\
In GR, the interest in the study of spacetime symmetries is long-standing. A large body of the literature is devoted to the classification of spacetimes according to different symmetries for they describe conservation laws such as conservation of energy, linear momentum and angular momentum in the form of first integrals of a dynamical system \cite{[5],[6]}. \\
Particularly, since $R_{ab}=R^c_{acb}$ where $R^c_{acb}$ denotes the Riemann curvature tensor, the symmetries of Ricci tensor have a natural geometric significance \cite{[7],[8],[9]}. As far their physical implications are concerned, Davis et al. \cite{[10],[11]} discussed this aspect in detail via the conservation laws admitted by particular types of matter fields. They also discussed some interesting applications of the results to relativistic hydrodynamics and plasma physics. Using RCs, Oliver and Davis \cite{[12]} obtained conservation expressions for perfect fluids. Tsamparlis and Mason \cite{[13]} studied certain properties of fluid spacetimes admitting RCs and a variety of imperfect fluids with cosmological constant and with anisotropic pressure. Apart from these physical implications, RCs can also be used to find the new exact solutions and in the classification of known solutions of Einstein's Field Equations (EFEs) \cite{[14],[15],[16],[17],[18],[19],[20],[21],[22],[23]}.\\
As far conformal symmetries are concerned, they are of particular interest as they provide a deeper insight into the spacetime geometry and are physically significant. The CKVs map null geodesics to null geodesics and preserve the structure of null cone. Also, they generate constants of motion along null geodesics for massless particles. This property connects the conformal symmetry with a well defined, physically meaningful conserved quantity. The effects of conformal symmetries can also be seen at kinematic and dynamic levels \cite{[24]}.\\
After the spacetime metric, Ricci and energy-momentum tensors are important candidates as they play a significant role in relativity towards understanding the geometric and physical structures of spacetimes. Whereas the conformal symmetries are extensively studied in literature \cite{[25],[26],[27],[28],[29],[30]},  the Ricci and energy-momentum tensors have not been so widely considered from the view point of these symmetries \cite{[4],[31]}. Keeping in mind the fact that classification of spacetimes by conformal collineations of Ricci and energy tensors would be of interest, we provide a complete classification of static plane symmetric spacetimes via CRCs and CMCs in this paper. The plan of the paper is as follows:\\
In next section, we present ten coupled CRC and CMC equations for static plane symmetric spacetimes. In section 3, we present a classification of these spacetimes according to their CRCs and CMCs when Ricci and energy momentum tensors are non-degenerate, while in section 4, the CRC and CMC equations are solved for degenerate case. The exact form of some static plane symmetric spacetime metrics admitting non-trivial CRCs and CMCs are provided in section 5. In the last section, we conclude our work with a brief summary and some physical implications of the obtained results.
\section{CRC and CMC Equations}
Plane symmetric spacetimes are Lorentzian manifolds admitting 3-dimensional isometry group $G_3$ in such a way that the group orbits are spacelike surfaces
of constant curvature. In static case, these spacetimes admit an additional timelike Killing vector. The most general form of line element for these spacetimes is given by \cite{[32]}.
\begin{equation}
ds^2=e^{2\alpha(x)} \ dt^2-dx^2-e^{2\beta(x)}\left[dy^2+dz^2\right], \label{eq:(2.1)}
\end{equation}
where $\alpha$ and $\beta$ are functions of $x$ only. Following are the four KVs admitted by the above metric:
\begin{equation}
\xi_{(1)}=\partial_t,\  \xi_{(2)}=\partial_y,\ \xi_{(3)}=\partial_z, \ \xi_{(4)}=z\partial_y-y\partial_z.    \label{eq:(2.2)}
\end{equation}
The surviving components of the Ricci tensor for the metric (\ref{eq:(2.1)}) are:
\begin{eqnarray}
R_{00}& = &e^{2\alpha} \left(\alpha''+2\alpha'\beta'+\alpha'^2\right)=A(x),  \nonumber \\
R_{11}& = & -2\left(\beta''+\beta'^2\right)-\left(\alpha''+\alpha'^2\right)=B(x) ,  \nonumber \\
R_{22}& = & R_{33}= -e^{2\beta} \left(\beta''+\alpha'\beta'+2\beta'^2\right)=C(x), \label{eq:(2.3)}
\end{eqnarray}
where the primes on the metric functions denote their derivative with respect to $x$. The Ricci scalar $R$ is found as:
\begin{equation}
R=2(\alpha''+\alpha'^2)+4(\alpha'\beta'+\beta'')+6 \beta'^2.    \label{eq:(2.4)}
\end{equation}
Using the Einstein's field equations (EFEs), $R_{ab}-\frac{R}{2}\ g_{ab}=k T_{ab},$ the energy-momentum tensor components are obtained as:
\begin{eqnarray}
T_{00}& = & - e^{2\alpha} \left(2\beta''+3\beta'^2\right) ,  \nonumber \\
T_{11}& = & \beta'^2+2\alpha' \beta' ,  \nonumber \\
T_{22}& = & T_{33}= e^{2\beta} \left(\beta''+\alpha'\beta'+\beta'^2+\alpha''+\alpha'^2\right), \label{eq:(2.5)}
\end{eqnarray}
Using the above Ricci tensor components in Eq. (\ref{eq:(1.2)}), we have the following CRC equations:
\begin{eqnarray}
A' \ \xi^1+2A\  \xi^0_{,0}& = &2\lambda A, \label{eq:(2.6)} \\
A \ \xi^0_{,1}+B\ \xi^1_{,0}& = & 0 ,  \label{eq:(2.7)} \\
A \ \xi^0_{,2}+C\  \xi^2_{,0}& = & 0 , \label{eq:(2.8)} \\
A \ \xi^0_{,3}+C\  \xi^3_{,0}& = & 0 , \label{eq:(2.9)} \\
B'\  \xi^1+2B \ \xi^1_{,1} & = &2\lambda B, \label{eq:(2.10)} \\
B \ \xi^1_{,2}+ C \ \xi^2_{,1} & = & 0 , \label{eq:(2.11)} \\
B \ \xi^1_{,3}+ C \ \xi^3_{,1} & = & 0 , \label{eq:(2.12)} \\
C' \ \xi^1+ 2C \ \xi^2_{,2} & = & 2\lambda C, \label{eq:(2.13)} \\
C\ \left(\xi^2_{,3}+\xi^3_{,2}\right)& = & 0 , \label{eq:(2.14)} \\
C'\  \xi^1+ 2C\  \xi^3_{,3} & = & 2\lambda C. \label{eq:(2.15)}
\end{eqnarray}
One can obtain the CMC equations by replacing $A(x)=T_{00},$  $B(x)=T_{11}$ and $C(x)=T_{22}=T_{33}$ in the above system.  To find the explicit form of CRCs and CMCs in static plane symmetric spacetimes, we need to solve the above set of equations. In the forthcoming sections, we solve these equations only for the CRCs. The similar procedure can be used to get CMCs.
\section{CRCs for Non-degenerate Ricci Tensor}
In this section, we solve Eqs. (\ref{eq:(2.6)})-(\ref{eq:(2.15)}) by considering the Ricci tensor $R_{ab}$ to be non-degenerate, that is $det R_{ab}\neq 0.$ This means that $A,\ B$ and $C$ all are non zero. \\
We differentiate Eqs. (\ref{eq:(2.8)}), (\ref{eq:(2.11)}) with respect to $z,$ Eqs. (\ref{eq:(2.9)}), (\ref{eq:(2.12)}) with respect to $y$ and Eq. (\ref{eq:(2.14)}) with respect to $t$ and $x$ to get the the following identities:
\begin{equation}
\xi^2_{,03}=\xi^0_{,23}=\xi^1_{,23}=\xi^2_{,13}=\xi^3_{,02}=\xi^3_{,12}=0. \label{eq:(3.1)}
\end{equation}
Dividing Eqs. (\ref{eq:(2.10)}) and (\ref{eq:(2.13)}) by $2B$ and $2C$ respectively, subtracting the resulting equations and then differentiating it with respect to $y$ and $z,$ we find that $\xi^2_{,223}=0.$ Thus we have:
\begin{equation}
\xi^2_{,3}=y\ F^1(t,x,z)+F^2(t,x,z),  \label{eq:(3.2)}
\end{equation}
where $F^1(t,x,z)$ and $F^2(t,x,z)$ are functions of integration. We use this value of $\xi^2_{,3}$ in Eq. (\ref{eq:(2.14)}) and integrate it with respect to $y$ to obtain:
\begin{equation}
\xi^3=-\frac{y^2}{2}\  F^1(t,x,z)-y\ F^2(t,x,z)+F^3(t,x,z),   \label{eq:(3.3)}
\end{equation}
$F^3(t,x,z)$ being a function of integration. If we subtract Eq. (\ref{eq:(2.15)}) from Eq. (\ref{eq:(2.13)}), we get:
\begin{equation}
\xi^2_{,2}=\xi^3_{,3} \ .  \label{eq:(3.4)}
\end{equation}
Differentiating the above equation once with respect to $z$ and twice with respect to $y,$ we find $F^1(t,x,z)=z\ f^1(t,x)+f^2(t,x),$ $F^2(t,x,z)=z\ f^3(t,x)+f^4(t,x)$ and $F^3(t,x,z)=\frac{z^3}{6}\ f^1(t,x)+\frac{z^2}{2}\ f^2(t,x)+z\ f^5(t,x)+f^6(t,x)$, where $f^i(t,x)$ denote function of integration for $i=1,...,6.$ We use these values of $F^2$ and $F^3$ in Eq. (\ref{eq:(3.2)}) and integrate it with respect to $z$ to obtain:
\begin{equation}
\xi^2=y\ \left(\frac{z^2}{2}\ f^1(t,x)+zf^2(t,x)\right)+\frac{z^2}{2}\ f^3(t,x)+zf^4(t,x)+F^4(t,x,y),  \label{eq:(3.5)}
\end{equation}
where $F^4(t,x,y)$ is a function of integration. Using the values of $\xi^2$ and $\xi^3$ from Eqs. (\ref{eq:(3.3)}) and (\ref{eq:(3.5)}) in Eq. (\ref{eq:(3.4)}) and then integrating it with respect to $y,$ we have $F^4(t,x,y)=-\frac{y^3}{6}\ f^1(t,x)-\frac{y^2}{2}\ f^3(t,x)+y\ f^5(t,x)+f^7(t,x),$ $f^7(t,x)$ being a function of integration. Finally, the two components $\xi^2$ and $\xi^3$ given in (\ref{eq:(3.3)}) and (\ref{eq:(3.5)}) take the form:
\begin{eqnarray*}
\xi^2&=&y\ \left(\frac{z^2}{2}\ f^1(t,x)+zf^2(t,x)\right)+\frac{z^2}{2}\ f^3(t,x)+zf^4(t,x)-\frac{y^3}{6}\ f^1(t,x) \nonumber \\
&-&\frac{y^2}{2}\ f^3(t,x)+y\ f^5(t,x)+f^7(t,x),
\end{eqnarray*}
\begin{eqnarray}
\xi^3&=&-\frac{y^2}{2}\  \biggl(z\ f^1(t,x)+f^2(t,x)\biggr)-y\ \biggl(z\ f^3(t,x)+f^4(t,x)\biggr)+\frac{z^3}{6}\ f^1(t,x) \nonumber \\
&+&\frac{z^2}{2}\ f^2(t,x)+z\ f^5(t,x)+f^6(t,x)   \label{eq:(3.6)}
\end{eqnarray}
Using the above value of $\xi^2$ in Eq. (\ref{eq:(2.8)}) and integrating it with respect to $y,$ we have:
\begin{eqnarray}
\xi^0&=&- \frac{C}{A}\ \biggl[ \frac{y^2}{2}\ \left(\frac{z^2}{2}\ f^1_t(t,x)+zf^2_t(t,x)\right)+y\left(\frac{z^2}{2}\ f^3_t(t,x)+zf^4_t(t,x)\right)-\frac{y^4}{24}\ f^1_t(t,x) \nonumber \\
&-&\frac{y^3}{6}\ f^3_t(t,x)+\frac{y^2}{2}\ f^5_t(t,x)+yf^7_t(t,x) \biggr]+F^5(t,x,z),   \label{eq:(3.7)}
\end{eqnarray}
where $F^5(t,x,z)$ is a function of integration. Using the above values of $\xi^0$ and $\xi^3$ in Eq. (\ref{eq:(2.9)}) and doing some simple algebraic manipulation, one gets $f^1_t(t,x)=f^2_t(t,x)=f^3_t(t,x)=f^4_t(t,x)=0$ and $F^5(t,x,z)=- \frac{C}{A}\ \left\{ \frac{z^2}{2}\ f^5_t(t,x)+zf^6_t(t,x)\right\}+f^8(t,x)$. With these values, (\ref{eq:(3.7)}) becomes:
\begin{eqnarray}
\xi^0&=&- \frac{C}{A}\ \biggl[\frac{y^2}{2}\ f^5_t(t,x)+yf^7_t(t,x)+\frac{z^2}{2}\ f^5_t(t,x)+zf^6_t(t,x) \biggr]+f^8(t,x),   \label{eq:(3.8)}
\end{eqnarray}
$f^8(t,x)$ being a function of integration. Similarly putting the value of $\xi^2$ from (\ref{eq:(3.6)}) in Eq. (\ref{eq:(2.11)}) and some straightforward calculation yields $f^1_x(t,x)=f^2_x(t,x)=f^3_x(t,x)=f^4_x(t,x)=0$ and
\begin{eqnarray}
\xi^1&=&- \frac{C}{B}\ \biggl[\frac{y^2}{2}\ f^5_x(t,x)+yf^7_x(t,x)+\frac{z^2}{2}\ f^5_x(t,x)+zf^6_x(t,x) \biggr]+f^9(t,x),   \label{eq:(3.9)}
\end{eqnarray}
where $f^9(t,x)$ denotes a function which arises in the process of integration. Thus $f^i(t,x)=a_i$, where $a_i$ is a constant for $i=1,...,4$ and we have the following components of the vector field generating CRCs, in terms of some unknown functions of $t$ and $x$.
\begin{eqnarray}
\xi^0&=&- \frac{C}{A}\ \biggl[\frac{y^2+z^2}{2}\ f^5_t(t,x)+y\ f^7_t(t,x)+z\ f^6_t(t,x) \biggr]+f^8(t,x), \nonumber \\
\xi^1&=&- \frac{C}{B}\ \biggl[\frac{y^2+z^2}{2}\ f^5_x(t,x)+y\ f^7_x(t,x)+z\ f^6_x(t,x) \biggr]+f^9(t,x), \nonumber \\
\xi^2&=& a_1 \biggl( \frac{y z^2}{2}-\frac{y^3}{6}\biggr)+a_2 yz+a_3 \biggl( \frac{z^2-y^2}{2}\biggr)+a_4z+y\ f^5(t,x)+f^7(t,x), \nonumber \\
\xi^3&=& a_1 \biggl( \frac{ z^3}{6}-\frac{y^2 z}{2}\biggr)+a_2 \biggl( \frac{z^2-y^2}{2}\biggr)-a_3 yz-a_4 y+z\ f^5(t,x)+f^6(t,x). \label{eq:(3.10)}
\end{eqnarray}
Using Eq. (\ref{eq:(2.13)}), the conformal factor $\lambda$ can be written as:
\begin{eqnarray}
\lambda &=& -\frac{C'}{2B}\ \biggl[\frac{y^2+z^2}{2}\ f^5_x(t,x)+y\ f^7_x(t,x)+z\ f^6_x(t,x) \biggr]+\frac{C'}{2C}\ f^9(t,x) \nonumber \\
&+& a_1 \biggl( \frac{z^2-y^2}{2}\biggr)+a_2 z-a_3 y+f^5(t,x).  \label{eq:(3.11)}
\end{eqnarray}
Substituting $\xi^a$ from (\ref{eq:(3.10)}) and $\lambda$ from (\ref{eq:(3.11)}) in Eqs. (\ref{eq:(2.6)}), (\ref{eq:(2.7)}) and (\ref{eq:(2.10)}), we have $a_1=0$ and the following integrability conditions are generated:
\begin{eqnarray}
\frac{1}{2}\biggl(\frac{C'}{C}-\frac{A'}{A} \biggr)\  f^5_x(t,x)- \frac{B}{A}\ f^5_{tt}(t,x)&=&0 \label{eq:(3.12)} \\
\frac{1}{2}\biggl(\frac{C'}{C}-\frac{A'}{A} \biggr)\  f^6_x(t,x)- \frac{B}{A}\ f^6_{tt}(t,x)-a_2\ \frac{B}{C}&=&0 \label{eq:(3.13)} \\
\frac{1}{2}\biggl(\frac{C'}{C}-\frac{A'}{A} \biggr)\  f^7_x(t,x)- \frac{B}{A}\ f^7_{tt}(t,x)+a_3\ \frac{B}{C}&=&0 \label{eq:(3.14)} \\
2C\ f^5_{tx}(t,x)+A\biggl(\frac{C}{A} \biggr)'\ f^5_t(t,x)&=&0 \label{eq:(3.15)} \\
2C\ f^6_{tx}(t,x)+A\biggl(\frac{C}{A} \biggr)'\ f^6_t(t,x)&=&0 \label{eq:(3.16)} \\
2C\ f^7_{tx}(t,x)+A\biggl(\frac{C}{A} \biggr)'\ f^7_t(t,x)&=&0 \label{eq:(3.17)} \\
\frac{C}{B}\ f^5_{xx}(t,x)+\frac{1}{2}\biggl(\frac{C}{B} \biggr)'\ f^5_x(t,x)&=&0 \label{eq:(3.18)} \\
\frac{C}{B}\ f^6_{xx}(t,x)+\frac{1}{2}\biggl(\frac{C}{B} \biggr)'\ f^6_x(t,x)+a_2&=&0 \label{eq:(3.19)} \\
\frac{C}{B}\ f^7_{xx}(t,x)+\frac{1}{2}\biggl(\frac{C}{B} \biggr)'\ f^7_x(t,x)-a_3&=&0 \label{eq:(3.20)} \\
\frac{1}{2}\biggl(\frac{A'}{A}-\frac{C'}{C} \biggr)\  f^9(t,x)+f^8_t(t,x)-f^5(t,x) &=&0 \label{eq:(3.21)}
\end{eqnarray}
\begin{eqnarray}
\frac{1}{2}\biggl(\frac{B'}{B}-\frac{C'}{C} \biggr)\  f^9(t,x)+f^9_x(t,x)-f^5(t,x) &=&0 \label{eq:(3.22)} \\
A\ f^8_x(t,x)+B\ f^9_t(t,x)&=&0 \label{eq:(3.23)}
\end{eqnarray}
The CRCs and the associated conformal factor given in (\ref{eq:(3.10)}) and (\ref{eq:(3.11)}) will get their final form if we could solve the system of Eqs. (\ref{eq:(3.12)})-(\ref{eq:(3.23)}). To write the CRCs, conformal factor and the integrability conditions in a more compact form, here we introduce new variables $\eta_i=\left(\eta_1,\eta_2,\eta_3\right)=\left(\frac{y^2+z^2}{2}, z, y\right)$, $P^i=\left(P^1,P^2,P^3\right)=\left(f^5,f^6,f^7\right)$ and denote $P^0=f^8$, $P^4=f^9.$ With these variables and keeping the Einstein summation convention in mind, one can write the system (\ref{eq:(3.10)}) and the conformal factor given in (\ref{eq:(3.11)}) as follows:
\begin{eqnarray}
\xi^0&=&- \frac{C}{A}\ \eta_i\ P^i_t+P^0, \nonumber \\
\xi^1&=&- \frac{C}{B}\ \eta_i\ P^i_x+P^4, \nonumber \\
\xi^2&=& (\eta_i)_{,2}\ P^i+a_2 yz+a_3 \biggl( \frac{z^2-y^2}{2}\biggr)+a_4z, \nonumber \\
\xi^3&=& (\eta_i)_{,3}\ P^i+a_2 \biggl( \frac{z^2-y^2}{2}\biggr)-a_3 yz-a_4 y, \nonumber \\
\lambda &=& -\frac{C'}{2B}\ \eta_i\ P^i_x+\frac{C'}{2C}\ P^4+P^1+a_2 z-a_3 y. \label{eq:(3.24)}
\end{eqnarray}
Also, the integrability conditions (\ref{eq:(3.12)})-(\ref{eq:(3.23)}) reduce to:
\begin{eqnarray}
\frac{1}{2}\biggl(\frac{C'}{C}-\frac{A'}{A} \biggr)\  P^i_x(t,x)- \frac{B}{A}\ P^i_{tt}(t,x)+k_i\ \frac{B}{C}&=&0, \label{eq:(3.25)} \\
2C\ P^i_{tx}(t,x)+A\biggl(\frac{C}{A} \biggr)'\ P^i_t(t,x)&=&0, \label{eq:(3.26)} \\
\frac{C}{B}\ P^i_{xx}(t,x)+\frac{1}{2}\biggl(\frac{C}{B} \biggr)'\ P^i_x(t,x)-k_i&=&0, \label{eq:(3.27)} \\
\frac{1}{2}\biggl(\frac{A'}{A}-\frac{C'}{C} \biggr)\  P^4(t,x)+P^0_t(t,x)-P^1(t,x) &=&0, \label{eq:(3.28)} \\
\frac{1}{2}\biggl(\frac{B'}{B}-\frac{C'}{C} \biggr)\  P^4(t,x)+P^4_x(t,x)-P^1(t,x) &=&0, \label{eq:(3.29)} \\
A\ P^0_x(t,x)+B\ P^4_t(t,x)&=&0, \label{eq:(3.30)}
\end{eqnarray}
where $k_i=0,-a_2,a_3$ for $i=1,2,3$ respectively. In order to have a complete classification in non-degenerate Ricci tensor case, we solve Eqs. (\ref{eq:(3.25)})-(\ref{eq:(3.30)}) by considering the following cases: \\
\ \ \ \ \ \ \ \  \textbf{(ND1)}\ $A'=B'=C'=0$ \qquad \ \ \ \ \ \ \ \ \ \ \ \ \ \ \ \  \textbf{(ND2)} $B'=C'=0$,\ $A'\neq0$\\
\ \ \ \ \  \ \ \ \textbf{(ND3)}\ $A'=C'=0$,\ $B'\neq0$ \qquad \ \ \ \ \ \ \ \ \ \ \ \ \   \textbf{(ND4)} $A'=B'=0$,\ $C'\neq0$ \\
\ \ \ \ \ \ \ \ \ \textbf{(ND5)}\ $A'=0$,\ $B'\neq0,$\  $C'\neq0$ \qquad \ \ \ \ \ \ \ \ \ \  \textbf{(ND6)} $A'\neq0$,\ $B'=0,$\  $C'\neq0$ \\
\ \ \ \ \ \ \ \ \ \textbf{(ND7)}\ $A'\neq0$,\ $B'\neq0,$\  $C'=0$ \qquad \ \ \ \ \ \ \ \ \ \  \textbf{(ND8)} $A'\neq0$,\ $B'\neq0,$\  $C'\neq0$ \\
It is well known that every KV is a RC, so the set of CRCs obtained by solving the system of Eqs. (\ref{eq:(3.25)})-(\ref{eq:(3.30)}) for each of the above case contain the four basic KVs of these spacetimes, mentioned in (\ref{eq:(2.2)}). In Tables 1-8 (Appendix A), we summarize the obtained results for all the above cases by presenting the CRCs, the corresponding conformal factor and the constraints on the Ricci tensor components under which static plane symmetric spacetimes admit these CRCs. Note that in some cases, when the derivative of the Ricci tensor components $A, B, C$ vanishes, the corresponding Ricci tensor component is assumed to be one.\\
Moreover, because of the mathematical similarity between CRC and CMC equations, we can directly write the explicit form of CMCs without solving CMC equations. Certainly, the same process will be repeated by replacing $A,\ B$ and $C$ by $T_{00},\ T_{11}$ and $T_{22}$ respectively and we get the same expressions for the explicit form of CMCs as those of CRCs with the only difference in the constraints to be satisfied by the energy-momentum tensor components.
\section{CRCs for Degenerate Ricci Tensor}
For the degenerate Ricci tensor, we have $det R_{ab}=ABC^2=0$ which give rise to the following possibilities: \\
%\ \ \ \ \ \ \ \  \textbf{(D1)} $A\neq0$,\ $B=C=0$ \qquad \ \ \ \ \ \ \ \ \ \ \ \ \  \textbf{(D2)} $A=C=0$,\ $B\neq0$, \\
%\ \ \ \ \  \ \ \ \textbf{(D3)}\ $A=B=0$,\ $C\neq0$ \qquad \ \ \ \ \ \ \ \ \ \ \ \ \   \textbf{(D4)} $A\neq0$,\ $B=0,$\  $C\neq0$ \\
%\ \ \ \ \ \ \ \ \ \textbf{(D5)}\ $A=0$,\ $B\neq0,$\  $C\neq0$ \qquad \ \ \ \ \ \ \ \ \ \  \textbf{(D6)} $A\neq0$,\ $B\neq0,$\  $C=0$ \\
\ \ \ \ \ \ \ \  \textbf{(D1)} $A\neq0$,\ $B\neq0,$\  $C=0$ \qquad \ \ \ \ \ \ \ \ \ \ \ \ \  \textbf{(D2)} $A\neq0$,\ $B=0,$\  $C\neq0$, \\
\ \ \ \ \  \ \ \ \textbf{(D3)}\ $A=0$,\ $B\neq0,$\  $C\neq0$ \qquad \ \ \ \ \ \ \ \ \ \ \ \ \   \textbf{(D4)} $A\neq0$,\ $B=C=0$ \\
\ \ \ \ \ \ \ \ \ \textbf{(D5)}\ $A=C=0$,\ $B\neq0$ \qquad \ \ \ \ \ \ \ \ \ \ \ \ \ \ \ \  \textbf{(D6)} $A=B=0$,\ $C\neq0$ \\
In each of the above cases, we directly solve the system of Eqs. (\ref{eq:(2.6)})-(\ref{eq:(2.15)}). In the first case (i.e. \textbf{D1}), we are left with the following equations:
\begin{eqnarray}
A' \ \xi^1+2A\  \xi^0_{,0}& = &2\lambda A, \label{eq:(4.1)} \\
A \ \xi^0_{,1}+B\ \xi^1_{,0}& = & 0 ,  \label{eq:(4.2)} \\
\xi^0_{,2} = \xi^0_{,3}=\xi^1_{,2}=\xi^1_{,3}&=&0  , \label{eq:(4.3)} \\
B'\  \xi^1+2B \ \xi^1_{,1} & = &2\lambda B. \label{eq:(4.4)}
\end{eqnarray}
Solving Eqs. (\ref{eq:(4.2)}) and (\ref{eq:(4.3)}), we have
\begin{eqnarray}
\xi^0&=&f_t(t,x), \nonumber \\
\xi^1&=& - \frac{A}{B}\ f_x(t,x)+g(x), \label{eq:(4.5)}
\end{eqnarray}
where $f(t,x)$ and $g(x)$ are unknown functions of integration. The conformal factor $\lambda$ gets the following form by using (\ref{eq:(4.5)}) in Eq. (\ref{eq:(4.4)}):
\begin{equation}
\lambda=\left(\frac{AB'-2A'B}{2B^2}\right)\ f_x(t,x)-\frac{A}{B}\ f_{xx}(t,x)+\frac{B'}{2B}\ g(x)+g_x(x). \label{eq:(4.6)}
\end{equation}
Substituting the values from Eqs. (\ref{eq:(4.5)}) and (\ref{eq:(4.6)}) in Eq. (\ref{eq:(4.1)}), we have:
\begin{equation}
\frac{1}{2} \left(\frac{A}{B}\right)'\ f_x(t,x)+\frac{A}{B}\ f_{xx}(t,x)+f_{tt}(t,x)+\left(\frac{A'}{2A}-\frac{B'}{2B}\right)\ g(x)-g_x(x)=0. \label{eq:(4.7)}
\end{equation}
The above equation is highly non linear and cannot be solved generally. However, one can choose some specific values of the functions $f(t,x)$ and $g(x)$ such that it holds true. With these values of $f(t,x)$ and $g(x),$ we would be able to write the two components $\xi^0$ and $\xi^1$ of CRCs presented in (\ref{eq:(4.5)}) in the final form. The other two components $\xi^2$ and $\xi^3$ are arbitrary functions of $t,x,y$ and $z.$ Hence there are infinite number of CRCs in this case, subject to the differential constraint given in Eq. (\ref{eq:(4.7)}).\\
It is straightforward to solve the CRC Eqs. (\ref{eq:(2.6)})-(\ref{eq:(2.15)}) in the remaining five cases and to see that each case yields infinite number of CRCs. We exclude the basic calculations and present the obtained results of all these cases (i.e. \textbf{D2 - D6}) in Table 9 (Appendix B).\\
Similarly, we can solve the CMC equations for degenerate energy-momentum tensor to deduce that there are infinitely many CMCs. Like the non-degenerate case, the explicit form of CMCs will be same as those of CRCs with the only difference that $A, B$ and $C$ are replaced by $T_{00}, T_{11}$ and $T_{22}$ which yield different differential constraints as those in case of CRCs.
\section{Spacetimes Admitting Non-Trivial CRCs and CMCs}
From tables 1-9, we see that the static plane symmetric spacetimes admit 7, 10 or 15 CRCs and CMCs for non-degenerate Ricci and energy-momentum tensors. However, the CRCs and CMCs have infinite degrees of freedom when the Ricci and energy-momentum tensors are degenerate. In order to show that the classes of CRCs and CMCs are non-empty, we need to construct some specific plane symmetric metrics satisfying the constraints given in each case. In this section, we give some examples of such metrics.
\begin{exmp}
Consider the following static plane symmetric metric admitting seven RCs \cite{[21]}:
\begin{equation}
ds^2=x^2\ dt^2-dx^2-x^4\ \left(dy^2+dz^2\right), \label{eq:(5.1)}
\end{equation}
The above metric satisfies all the constraints of case \textbf{ND5} for $k=-\frac{1}{4}$ and hence it admits 15 CRCs as given in Table 4. Out of these 15 CRCs, seven are RCs and remaining eight are proper CRCs.  The energy-momentum tensor components for the above metric are $T_{00}=-8, T_{11}=\frac{8}{x^2}$ and $T_{22}=T_{33}=4x^2,$ which satisfy the constraints of case \textbf{ND5} for $k=\frac{1}{8}$, if we replace $A, B$ and $C$ by  $T_{00}, T_{11}$ and $T_{22}$ respectively. Hence there are 15 CMCs in this case, out of which seven are MCs and the remaining eight are proper CMCs. We can easily check that the metric (\ref{eq:(5.1)}) is conformally flat.
\end{exmp}
\begin{exmp}
Here we consider the following static plane symmetric metric:
\begin{equation}
ds^2= dt^2-dx^2-e^x\ \left(dy^2+dz^2\right). \label{eq:(5.2)}
\end{equation}
The energy-momentum tensor components for the above metric are $T_{00}=-\frac{3}{4}, T_{11}=\frac{1}{4}$ and $T_{22}=T_{33}=\frac{e^x}{4}.$ These values clearly satisfy the constraints of case \textbf{ND4(c)} for $ \kappa=\frac{1}{4},$ if we replace $A, B$ and $C$ by  $T_{00}, T_{11}$ and $T_{22}$ respectively. Thus we have fifteen CMCs as presented in table 3. Out of these fifteen CMCs, seven are MCs and the other eight are proper CMCs. The Ricci tensor components for the metric  (\ref{eq:(5.2)}) become $A=0, B=-\frac{1}{2}$ and $C=-\frac{e^x}{2},$ giving the constraints of degenerate Ricci tensor case \textbf{D3}. Thus we have infinitely many CRCs which are presented in table 9.
\end{exmp}
\begin{exmp}
Here we take the following static plane symmetric metric:
\begin{equation}
ds^2= x^{\frac{4}{3}}\ \left(dt^2-dy^2-dz^2\right)-dx^2. \label{eq:(5.3)}
\end{equation}
The energy-momentum tensor components for this metric are $T_{00}=T_{22}=T_{33}=0$ and $T_{11}=\frac{12}{9x^2},$ which satisfy the constraints of case \textbf{D5}. Thus we have infinite CMCs for the above metric. The Ricci tensor components for the same metric are $A(x)=\frac{14}{9} x^{-\frac{2}{3}},$ $B(x)=-\frac{2}{9x^2}$ and $C(x)=-\frac{4}{9} x^{-\frac{2}{3}}.$ These $R_{ab}$'s satisfy the constraints of case \textbf{ND8(a)} for $k=-\frac{7}{2}$ and there are 15 CRCs as presented in table 7.
\begin{exmp}
For $\alpha=const.$ and $\beta=\frac{2}{3}\ln x,$ the static plane symmetric metric becomes:
\begin{equation}
ds^2= dt^2-dx^2-x^{\frac{4}{3}}\ \left(dy^2+dz^2\right). \label{eq:(5.4)}
\end{equation}
For this metric, we have $A(x)=0,$ $B(x)=\frac{4}{9x^2}$ and $C(x)=-\frac{2}{9} x^{-\frac{2}{3}}$ which satisfy the conditions of case \textbf{D3} and hence there are infinite number of CRCs. The Ricci scalar for the above metric vanishes and hence the energy-momentum tensor components coincide with the Ricci tensor components. So in this case we have infinite number of CMCs as well as CRCs.
\end{exmp}

\end{exmp}
\section{Summary and Discussion}
In this paper, we have presented a complete classification of static plane symmetric spacetimes according to their CRCs and CMCs by considering both degenerate and non-degenerate cases. It is the generalization of the previously published work on RCs by Farid et. al \cite{[21]}. From the present work, one can also obtain the classification of static plane symmetric spacetimes according to their HRCs and HMCs by considering the conformal function $\lambda$ to be a constant. Our study revealed that if the Ricci tensor is degenerate, then the CRCs have infinite degrees of freedom. It is worth noticing that the same spacetimes admit finite number of RCs when $A\neq0,\ C\neq0$ and $B=0$ \cite{[21]}, while in the same case we have obtained infinite number of CRCs (see case \textbf{D2}). Setting $\lambda=0$ in case \textbf{D2(a)}, it is easy to obtain the exactly ten RCs which are presented in \cite{[21]} under the same constraints. However, if we consider $\lambda=0$ in case \textbf{D2(b)}, we get the four basic KVs of static plane symmetric spacetimes if $C'\neq 0$ and infinite number of RCs if $C'=0.$ Under the same constraints, Farid et. al \cite{[21]} obtained one proper RC along with the four KVs, which appears wrong (see the system (2.2) therein). Similar remarks hold true in case of degenerate CMCs.\\
In case of non-degenerate Ricci or energy-momentum tensor, our classification shows that static plane symmetric spacetimes admit 7, 10 or 15-dimensional Lie algebra of CRCs or CMCs.  If we set $\lambda=0$ in all the non-degenerate cases, we can easily see that the number of independent RCs and MCs admitted by static plane symmetric spacetimes is 4,5,6,7 or 10 which is same as presented in \cite{[21]}.\\
Corresponding to every set of CRCs and CMCs presented here, there are some constraints to be satisfied by the Ricci and energy-momentum tensor components. The solution of these constraints would give the exact form of the static plane symmetric metric admitting these CRCs and CMCs. Due to the high non-linearity of the constraints, it is not easy to solve them generally. However, we have found some specific metrics satisfying these constraints.\\
Though the CRC and CMC equations appear to be similar, our classifications shows that their solutions may produce different Lie algebras of CRCs and CMCs. This fact can be seen in examples of section 5.\\
For the physical implications of our obtained results, we consider a perfect fluid as a source of the stress-energy tensor, $T_{ab}=(p+\rho) u_a u_b-p g_{ab},$ with $p,\rho$ and $u^a$ representing the pressure, energy density and four velocity of the perfect fluid respectively. We choose the fluid velocity as $u^a=e^{-\alpha} \delta^a_0,$ then we have $T_{00}=\rho e^{2\alpha},$ $T_{11}=p$ and $T_{22}=T_{33}=p e^{2\beta}.$ Thus we have the following perfect fluid matter tensor metric:
\begin{equation}
ds^2_{Perfect}=\rho e^{2\alpha}\ dt^2+p \ dx^2+p e^{2\beta}\ (dy^2+dz^2), \label{eq:(6.1)}
\end{equation}
which is positive definite if $\rho>0$ and $p>0.$ Using the above components of matter tensor in (\ref{eq:(2.5)}), we can write the Ricci components as:
\begin{eqnarray}
A(x)=e^{2\alpha}\biggl(\frac{3p+\rho}{2}\biggr),   B(x)= \frac{\rho-p}{2}, C(x)= e^{2\beta} \biggl(\frac{\rho-p}{2}\biggr). \label{eq:(6.2)}
\end{eqnarray}
The metric form of the Ricci tensor becomes:
\begin{equation}
2 ds^2_{Ric}=(3p+\rho)e^{2\alpha}\ dt^2+ (\rho-p)\ dx^2+e^{2\beta}(\rho-p)\ (dy^2+dz^2). \label{eq:(6.3)}
\end{equation}
The above Ricci tensor metric is positive definite if $\rho >0,$ $3p+\rho >0$ and $\rho > |p|.$ These conditions respectively represent weak, strong and dominant energy conditions. Following is the linear form of a barotropic equation of state $p=p(\rho)$ \cite{[16]}:
\begin{equation}
p=(\gamma -1) \rho,\label{eq:(6.4)}
\end{equation}
where $\gamma$ is a constant and correspond to dust filled universe, stiff matter and incoherent radiation for $\gamma=1, 2$ and $\frac{4}{3}$ respectively. Also $\gamma=0$ implies vacuum fluid. For a perfect fluid, the energy conditions are \cite{[16]}:
\begin{eqnarray}
\rho > 0, \ 0\leq p \leq \rho, \  0\leq \frac{dp}{d\rho} \leq 1.\label{eq:(6.5)}
\end{eqnarray}
From (\ref{eq:(6.2)}), it is clear that $B=0$ if and only if $C=0$. Similarly, for a perfect fluid we have $T_{11}=0$ if and only if $T_{22}=0.$ Thus a perfect fluid is not allowed in the degenerate cases \textbf{D1, D2, D5} and \textbf{D6}. The condition of a perfect fluid in case \textbf{D3} yields the equation of state $3p+\rho =0$ which violates the strong energy condition. Finally, the constraints in case \textbf{D4} implies $p=\rho=\frac{A}{2} e^{-2\alpha},$ which is the condition for stiff matter. Moreover, the constraints in case \textbf{D3} gives $\alpha''+2\alpha'\beta'+\alpha'^2=0.$ We may choose some specific metric functions $\alpha$ and $\beta$ satisfying this relation to form the exact form of perfect fluid static plane symmetric metric. A simplest example of such a perfect fluid metric is given in (\ref{eq:(5.2)}) where the pressure and energy density are given as $p=\frac{1}{4}$ and $\rho=-\frac{3}{4}$. Similarly, the conditions in case \textbf{D4} imply $2\left(\beta''+\beta'^2\right)+\left(\alpha''+\alpha'^2\right)=0$ and $\beta''+\alpha'\beta'+2\beta'^2=0.$ One may find some specific values of $\alpha$ and $\beta$ to get the exact form of such perfect fluid spacetime metric.\\
As far non-degenerate cases are concerned , perfect fluid is not allowed in the cases \textbf{ND1} and \textbf{ND2}, as for $B=C=const$ we must have $\beta=0\ \Rightarrow C=0,$ which is a contradiction to the condition of non-degenerate Ricci tensor. In the remaining non-degenerate cases, the conditions of perfect fluid establish relation between $p$ and $\rho,$ however the system  (\ref{eq:(2.3)}) yields highly non-linear equations involving the metric functions $\alpha$ and $\beta$ which need to be solved. One may work on these equations to get the exact form of the perfect fluid plane symmetric metrics admitting non-trivial CRCs and CMCs.

\begin{landscape}
{\Large \textbf{Appendix A}}
\begin{table}[h]
\centering
\footnotesize
%\label{table 1}
%\caption{\footnotesize CRCs for Non-Degenerate Ricci Tensor}
\begin{center}
\begin{tabular}{|l|c|c|c|}
\hline
 \ \ \ \ Case  \ \ \  & Constraints  &  CRCs \ \ \  & Conformal Factor ($\lambda$)\ \ \ \  \\
\hline
\hline
&&&\\
ND1 & --- & $\xi^0=\frac{c_2}{2}\ \left(t^2-x^2-y^2-z^2\right)+c_1 tx+c_3t-c_{10}x-c_4y-c_7z+a_2tz-a_3ty+c_{12},$ & $a_2z-a_3y+c_1x$ \\
 & & $\xi^1=\frac{c_1}{2}\ \left(x^2-t^2-y^2-z^2\right)+c_2 tx+c_{10}t+c_3x-c_6y-c_8z-a_3xy+a_2xz+c_{11},$ & $+c_2t+c_3$\\
 & & $\xi^2=\frac{a_3}{2}\ \left(x^2+t^2-y^2+z^2\right)+a_2 yz+c_1 xy+c_2 ty+a_4z+c_3y+c_4 t+c_6x+c_5,$ & \\
 & & $\xi^3=\frac{a_2}{2}\ \left(-t^2-x^2-y^2+z^2\right)-a_3 yz+c_1 xz+c_2 tz-a_4y+c_3z+c_7 t+c_8x+c_9.$& \\
&&&\\
\hline
&&&\\
ND2(a) & $A=\left(c_1x+c_2\right)^2$& $\xi^0=\frac{c_1}{\sqrt{A}} \biggl[ \frac{y^2+z^2}{2} \left(c_3\sin c_1t-c_4\cos c_1t\right)+y\left(c_5\sin c_1t-c_6\cos c_1t\right) $& $\sqrt{A} \left(c_3\cos c_1t+c_4\sin c_1t\right)$\\
 & & $+z \left(c_7\sin c_1t+c_8\cos c_1t\right)- \left(c_9\cos c_1t+c_{10}\sin c_1t\right)$& $+a_2z-a_3y+c_{12}$\\
 & & $+\frac{\left(A+c_2^2\right)}{2c_1^2}\left(c_3\sin c_1t-c_4\cos c_1t\right) \biggr]+c_{11},$&\\
 & & $\xi^1=-c_1 \biggl[ \frac{y^2+z^2}{2} \left(c_3\cos c_1t+c_4\sin c_1t\right)+y\left(c_5\cos c_1t+c_6\sin c_1t\right) $& \\
 & & $+z \left(c_7\cos c_1t+c_8\sin c_1t\right)+ c_9\sin c_1t-c_{10}\cos c_1t \biggr]$&\\
 & & $\frac{\sqrt{A}}{c_1} \left(a_2z-a_3y+c_{12}\right)+\frac{\left(A-c_2^2\right)}{2c_1}\left(c_3\cos c_1t+c_4\sin c_1t\right),$&\\
 & & $\xi^2=a_2yz+\frac{a_3}{2} (z^2-y^2)+a_4z+c_{12}y+y\sqrt{A} \left(c_3\cos c_1t+c_4\sin c_1t\right) $& \\
 & & $+\sqrt{A} \left(c_5\cos c_1t+c_6\sin c_1t\right)+\frac{a_3}{2c_1^2} \left(A-c_2^2\right)+c_{13}, $& \\
 & & $\xi^3=-a_3yz+\frac{a_2}{2} (z^2-y^2)-a_4y+c_{12}z+z\sqrt{A} \left(c_3\cos c_1t+c_4\sin c_1t\right) $& \\
 & & $+\sqrt{A} \left(c_7\cos c_1t+c_8\sin c_1t\right)-\frac{a_2}{2c_1^2} \left(A-c_2^2\right)+c_{14}. $& \\
&&&\\
\hline
\end{tabular}
\end{center}
\caption{\footnotesize CRCs for Non-Degenerate Ricci Tensor}
\end{table}
\end{landscape}

\begin{landscape}
\begin{table}[h]
\centering
\footnotesize
%\label{table 2}
%\caption{\footnotesize CRCs for Non-Degenerate Ricci Tensor}
\begin{center}
\begin{tabular}{|l|c|c|c|}
\hline
 \ \ \ \ Case  \ \ \  & Constraints  &  CRCs \ \ \  & Conformal Factor ($\lambda$)\ \ \ \  \\
\hline
\hline
ND2(b) & $A=\biggl( c_1x+c_2\biggr)^{2\left(1-\frac{c_3}{c_1}\right)},$      & $\xi^0=c_3t+c_4,\ \ \xi^1=c_1x+c_2,$               & $c_1$ \\
       & where $c_1\neq c_3 \neq 0$                                                 &  $\xi^2=a_4z+c_1y+c_5, \ \ \xi^3=-a_4y+c_1z+c_6.$  &       \\
&&&\\
\hline
&&&\\
ND3 & --- &  $\xi^0=\frac{c_1}{2}\ \left(t^2-y^2-z^2\right)-a_3 ty+a_2 tz-c_5y-c_3z+c_8t+\left(c_7t+c_2\right)\ \int \sqrt{B} dx$ &$a_2z-a_3y+c_1t$\\
    &     &  $-c_1 \int \left( \sqrt{B} \ \int \sqrt{B} dx \right) dx+c_4,$                                                    & $+c_7\int \sqrt{B} dx+c_8$ \\
    &     &  $\xi^1=-\frac{1}{\sqrt{B}} \biggl[ \frac{c_7}{2} \left( t^2+y^2+^2\right)+c_{11}y+c_9z+c_2t+\left(a_3y-a_2z-c_1t+c_8\right)\int \sqrt{B} dx$&\\
    &     &  $-c_7 \int \left( \sqrt{B} \ \int \sqrt{B} dx \right) dx-c_6 \bigg],$                                   &\\
    &     &  $\xi^2=\frac{a_3}{2}\ \left(t^2-y^2+z^2\right)+a_2 yz+c_1 ty+c_8y+a_4z+c_5t+\left(c_7y+c_{11}\right)\ \int \sqrt{B} dx$ & \\
    &     &  $+a_3 \int \left( \sqrt{B} \ \int \sqrt{B} dx \right) dx+c_{12},$                                                       & \\
    &     &  $\xi^3=\frac{a_2}{2}\ \left(-t^2-y^2+z^2\right)-a_3 yz+c_1 tz+c_8z-a_4y+c_3t+\left(c_7z+c_9\right)\ \int \sqrt{B} dx$ & \\
    &     &  $-a_2 \int \left( \sqrt{B} \ \int \sqrt{B} dx \right) dx+c_{10}.$                                                       & \\
&&&\\
\hline
&&&\\
ND4(a) & $C=\left(c_1x+c_2\right)^2$  & $\xi^0=\frac{c_1}{2}\ \left(t^2-x^2\right)+c_3t-c_2x+c_4$,\ \ \ \ $\xi^1=c_1xt+c_3x+c_2t+\frac{c_2c_3}{c_1}$ & $c_1t+c_3$ \\
 & & $\xi^2=a_4z+c_5,$ \ \ \ \ \ $\xi^3=-a_4y+c_6.$& \\
\hline
ND4(b) & $C=\biggl( c_1x+c_2\biggr)^{2\left(1-\frac{c_3}{c_1}\right)},$ & $\xi^0=c_1t+c_4,\ \ \xi^1=c_1x+c_5,$ & $c_1$ \\
       & where $c_1 \neq c_3 \neq 0$ &  $\xi^2=a_4z+c_3y+c_6, \ \ \xi^3=-a_4y+c_3z+c_7.$  &       \\
\hline
\end{tabular}
\end{center}
\caption{\footnotesize CRCs for Non-Degenerate Ricci Tensor}
\end{table}
\end{landscape}

\begin{landscape}
\begin{table}[h]
\centering
\footnotesize
%\label{table 3}
%\caption{\footnotesize CRCs for Non-Degenerate Ricci Tensor}
\begin{center}
\begin{tabular}{|l|c|c|c|}
\hline
 \ \ \ \ Case  \ \ \  & Constraints  &  CRCs \ \ \  & Conformal Factor ($\lambda$)\ \ \ \  \\
\hline
\hline
&&&\\
ND4(c) & $C=ke^x,$ & $\xi^0=-\sqrt{k} \ e^{\frac{x}{2}} \biggl[ \frac{y^2+z^2}{2} \left(c_8 \cos \frac{t}{2}+c_9 \sin \frac{t}{2}\right)+y\left(c_5 \cos \frac{t}{2}+c_6 \sin \frac{t}{2}\right)$ & $\frac{\sqrt{k}}{2} e^{\frac{x}{2}} \biggl[ \frac{y^2+z^2}{2} \left(c_8 \sin \frac{t}{2}-c_9 \cos \frac{t}{2}\right)$\\
 & where $k \in \mathbb{R}$  & $+z\left(c_1 \cos \frac{t}{2}+c_2 \sin \frac{t}{2}\right) \biggr]+e^{\frac{x}{2}} \left(c_{12} \cos \frac{t}{2}+c_{11} \sin \frac{t}{2}\right) $& $+y \left(c_5 \sin \frac{t}{2}-c_6 \cos \frac{t}{2}\right) $\\
 & & $-\frac{2}{\sqrt{k}} e^{\frac{-x}{2}} \left(c_8 \cos \frac{t}{2}+c_9 \sin \frac{t}{2}\right)+c_3,$& $+z \left(c_1 \sin \frac{t}{2}-c_2 \cos \frac{t}{2}\right) \biggr] $ \\
 & & $\xi^1=\sqrt{k} \ e^{\frac{x}{2}} \biggl[ \frac{y^2+z^2}{2} \left(c_8 \sin \frac{t}{2}-c_9 \cos \frac{t}{2}\right)+y\left(c_5 \sin \frac{t}{2}-c_6 \cos \frac{t}{2}\right)$ & $+\frac{1}{\sqrt{k}} e^{-\frac{x}{2}} \left(c_8 \sin \frac{t}{2}-c_9 \cos \frac{t}{2}\right) $\\
 &  & $+z\left(c_1 \sin \frac{t}{2}-c_2 \cos \frac{t}{2}\right) \biggr]-e^{\frac{x}{2}} \left(c_{12} \sin \frac{t}{2}-c_{11} \cos \frac{t}{2}\right) $& $-\frac{1}{2} e^{\frac{x}{2}} \left(c_{12} \sin \frac{t}{2}-c_{11} \cos \frac{t}{2}\right) $\\
 & & $-\frac{2}{\sqrt{k}} e^{\frac{-x}{2}} \left(c_8 \sin \frac{t}{2}-c_9 \cos \frac{t}{2}\right)+2a_3y-2a_2z-2c_{10},$&  \\
&&&\\
 & & $\xi^2=\frac{2y}{\sqrt{k}} e^{\frac{-x}{2}} \left(c_8 \sin \frac{t}{2}-c_9 \cos \frac{t}{2}\right)+\frac{2}{\sqrt{k}} e^{\frac{-x}{2}} \left(c_1 \sin \frac{t}{2}-c_2 \cos \frac{t}{2}\right)+\frac{2a_3}{\sqrt{k}} e^{-x}$& \\
 & & $+a_2yz+\frac{a_3}{2} \left(z^2-y^2\right)+a_4z+c_{10}y+c_7,$&\\
 &&&\\
  & & $\xi^3=\frac{2z}{\sqrt{k}} e^{\frac{-x}{2}} \left(c_8 \sin \frac{t}{2}-c_9 \cos \frac{t}{2}\right)+\frac{2}{\sqrt{k}} e^{\frac{-x}{2}} \left(c_1 \sin \frac{t}{2}-c_2 \cos \frac{t}{2}\right)-\frac{2a_2}{\sqrt{k}} e^{-x}$& \\
 & & $-a_3yz+\frac{a_2}{2} \left(z^2-y^2\right)-a_4y+c_{10}z+c_4.$&\\
&&&\\
\hline
\end{tabular}
\end{center}
\caption{\footnotesize CRCs for Non-Degenerate Ricci Tensor}
\end{table}
\end{landscape}

\begin{landscape}
\begin{table}[h]
\centering
\footnotesize
%\label{table 4}
%\caption{\footnotesize CRCs for Non-Degenerate Ricci Tensor}
\begin{center}
\begin{tabular}{|l|c|c|c|}
\hline
 \ \ \ \ Case  \ \ \  & Constraints  &  CRCs \ \ \  & Conformal Factor ($\lambda$)\ \ \ \  \\
\hline
\hline
&&&\\
ND5 & $B=\frac{C'^2}{4kC^2},$& $\xi^0=-\sqrt{C} \biggl[\frac{y^2+z^2}{2} \left(c_8 \sin \sqrt{k} t+c_9 \cos \sqrt{k} t\right)+y \left(c_5 \sin \sqrt{k} t+c_6 \cos \sqrt{k} t\right)$ & $-\sqrt{kC} \biggl[\frac{y^2+z^2}{2} \left(c_8 \cos \sqrt{k} t-c_9 \sin \sqrt{k} t\right)$ \\
 & where $k$ & $+z \left(c_1 \sin \sqrt{k} t+c_2 \cos \sqrt{k} t\right)-c_{12} \cos \sqrt{k} t-c_{11} \sin \sqrt{k} t \biggr]$ & $+y\left(c_5 \cos \sqrt{k} t-c_6 \sin \sqrt{k} t\right)$\\
 & is a constant & $-\frac{1}{2k \sqrt{C}} \left(c_8 \sin \sqrt{k} t+c_9 \cos \sqrt{k} t\right)+c_3,$& $+z\left(c_1 \cos \sqrt{k} t-c_2 \sin \sqrt{k} t\right)$ \\
 &&&\\
 & & $\xi^1=-\frac{2C \sqrt{k C}}{C'} \biggl[\frac{y^2+z^2}{2} \left(c_8 \cos \sqrt{k} t-c_9 \sin \sqrt{k} t\right)+y \left(c_5 \cos \sqrt{k} t-c_6 \sin \sqrt{k} t\right)$& $+\left(c_{12} \sin \sqrt{k} t-c_{11} \cos \sqrt{k} t\right) \biggr]$\\
 & & $+z \left(c_1 \cos \sqrt{k} t-c_2 \sin \sqrt{k} t\right)+c_{12} \sin \sqrt{k} t-c_{11} \cos \sqrt{k} t +\frac{1}{\sqrt{kC}} \left( a_2z-a_3y\right)\biggr]$& $-\frac{1}{2\sqrt{kC}}\left(c_8 \cos \sqrt{k} t-c_9 \sin \sqrt{k} t\right)$\\
 & & $+\frac{\sqrt{C}}{\sqrt{k}C'} \left(c_8 \cos \sqrt{k} t-c_9 \sin \sqrt{k} t\right)-\frac{2c_{10} C}{C'}, $& \\
 &&&\\
 & & $\xi^2=\frac{y}{\sqrt{kC}} \left(-c_8 \cos \sqrt{k} t+c_9 \sin \sqrt{k} t\right)-\frac{1}{\sqrt{kC}}\left(c_5 \cos \sqrt{k} t-c_6 \sin \sqrt{k} t\right) $& \\
 & & $+a_2yz+\frac{a_3}{2} \left( z^2-y^2\right)+a_4z+c_{10}y+\frac{a_3}{2kC}+c_7$& \\
&&&\\
 & & $\xi^3=\frac{z}{\sqrt{kC}} \left(-c_8 \cos \sqrt{k} t+c_9 \sin \sqrt{k} t\right)-\frac{1}{\sqrt{kC}}\left(c_1 \cos \sqrt{k} t-c_2 \sin \sqrt{k} t\right) $& \\
 & & $-a_3yz+\frac{a_2}{2} \left( z^2-y^2\right)-a_4y+c_{10}z-\frac{a_2}{2kC}+c_4$& \\
&&&\\
\hline
\end{tabular}
\end{center}
\caption{\footnotesize CRCs for Non-Degenerate Ricci Tensor}
\end{table}
\end{landscape}

\begin{landscape}
\begin{table}[h]
\centering
\footnotesize
%\label{table 5}
%\caption{\footnotesize CRCs for Non-Degenerate Ricci Tensor}
\begin{center}
\begin{tabular}{|l|c|c|c|}
\hline
 \ \ \ \ Case  \ \ \  & Constraints  &  CRCs \ \ \  & Conformal Factor ($\lambda$)\ \ \ \  \\
\hline
\hline
&&&\\
ND6 & $A=C\biggl( \int \frac{1}{\sqrt{C}} dx+k \biggr)^2$ & $\xi^0=\sqrt{\frac{C}{A}} \biggl[ \frac{y^2+z^2}{2} \left(c_1 \sin t-c_2 \cos t\right)+y\left(c_3 \sin t-c_4 \cos t\right) $ & $\frac{C'}{2\sqrt{C}}\biggl[ \frac{y^2+z^2}{2} \left(c_1 \cos t+c_2 \sin t\right)$\\
 & where $k$ is a constant.& $+z\left(c_5 \sin t-c_6 \cos t\right)+c_7 \cos t+c_8 \sin t \biggr]$ & $+y\left(c_3 \cos t+c_4 \sin t\right)$\\
 & & $+\frac{1}{2} \sqrt{\frac{A}{C}} \left(c_1 \sin t-c_2 \cos t\right)+c_9 $& $+z\left(c_5 \cos t+c_6 \sin t\right)$\\
 & & $\xi^1=-\sqrt{C} \biggl[ \frac{y^2+z^2}{2} \left(c_1 \cos t+c_2 \sin t\right)+y\left(c_3 \cos t+c_4 \sin t\right)$ & $-\frac{A}{2C} \left(c_1 \cos t+c_2 \sin t\right) $\\
 & & $+z\left(c_5 \cos t+c_6 \sin t\right)-\frac{A}{2C}\left(c_1 \cos t+c_2 \sin t\right)$& $-c_7\sin t+c_8\cos t $\\
 & & $-c_7 \sin t+c_8 \cos t+\sqrt{\frac{A}{C}} \left(a_3y-a_2z-c_{10}\right) \biggr ]$& $+\sqrt{\frac{A}{C}} \left(a_3y-a_2z-c_{10}\right) \biggr ]$\\
 & & $\xi^2=a_2yz+\frac{a_3}{2} \left(z^2-y^2\right)+a_4z+c_{10}y+\sqrt{\frac{A}{C}} y\left(c_1 \cos t+c_2 \sin t\right) $& $+\sqrt{\frac{A}{C}} \left(c_1 \cos t+c_2 \sin t\right)$\\
 & & $\sqrt{\frac{A}{C}} \left(c_3 \cos t+c_4 \sin t\right)+a_3 \int \biggl(\frac{1}{\sqrt{C}} \int \frac{1}{\sqrt{C}} dx \biggr) dx$& $+a_2z-a_3y+c_{10}$\\
 & & $+a_3 k \int \frac{1}{\sqrt{C}} dx+c_{11} $&\\
 & & $\xi^3=-a_3yz+\frac{a_2}{2} \left(z^2-y^2\right)-a_4y+c_{10}z+\sqrt{\frac{A}{C}} z\left(c_1 \cos t+c_2 \sin t\right) $& \\
 & & $\sqrt{\frac{A}{C}} \left(c_5 \cos t+c_6 \sin t\right)-a_2 \int \biggl(\frac{1}{\sqrt{C}} \int \frac{1}{\sqrt{C}} dx \biggr) dx$&\\
 & & $-a_2 k \int \frac{1}{\sqrt{C}} dx+c_{12} $&\\
&&&\\
\hline
\end{tabular}
\end{center}
\caption{\footnotesize CRCs for Non-Degenerate Ricci Tensor}
\end{table}
\end{landscape}

\begin{landscape}
\begin{table}[h]
\centering
\footnotesize
%\label{table 6}
%\caption{\footnotesize CRCs for Non-Degenerate Ricci Tensor}
\begin{center}
\begin{tabular}{|l|c|c|c|}
\hline
 \ \ \ \ Case  \ \ \  & Constraints  &  CRCs \ \ \  & Conformal Factor ($\lambda$)\ \ \ \  \\
\hline
\hline
&&&\\
ND7(a) & $B=\frac{A'^2}{A}$& $\xi^0=\frac{1}{2\sqrt{A}} \biggl[ \frac{y^2+z^2}{2} \left(c_1\sin\frac{t}{2}-c_2\cos\frac{t}{2}\right)+y\left(c_3\sin\frac{t}{2}-c_4\cos\frac{t}{2}\right)+z \left(c_5\sin\frac{t}{2}-c_6\cos\frac{t}{2}\right) \biggr]$& $\sqrt{A} \left(c_1\cos\frac{t}{2}+c_2\sin\frac{t}{2}\right)$\\
& &$+\sqrt{A}\left(c_1\sin\frac{t}{2}-c_2\cos\frac{t}{2}\right)+\frac{1}{\sqrt{A}} \left(c_7\cos\frac{t}{2}+c_8\sin\frac{t}{2}\right)+c_9$& $+a_2z-a_3y+c_{10}$\\
 &&&\\
 & &$\xi^1=-\frac{\sqrt{A}}{2A'}\biggl[ \frac{y^2+z^2}{2} \left(c_1\cos\frac{t}{2}+c_2\sin\frac{t}{2}\right)+y\left(c_3\cos\frac{t}{2}+c_4\sin\frac{t}{2}\right)+z \left(c_5\cos\frac{t}{2}+c_6\sin\frac{t}{2}\right) \biggr] $ &\\
 & & $+\frac{A\sqrt{A}}{A'} \left(c_1\cos\frac{t}{2}+c_2\sin\frac{t}{2}\right)+\frac{\sqrt{A}}{A'}\left(c_7\sin\frac{t}{2}-c_8\cos\frac{t}{2}\right)+\frac{2A}{A'} \left( a_2z-a_3y+c_{10}\right),$&\\
 &&&\\
 & & $\xi^2=a_2yz+\frac{a_3}{2} \left(z^2-y^2\right)+a_4z+c_{10}y+y\sqrt{A}\left(c_1\cos\frac{t}{2}+c_2\sin\frac{t}{2}\right) $&\\
 & & $+\sqrt{A}\left(c_3\cos\frac{t}{2}+c_4\sin\frac{t}{2}\right)+2a_3A+c_{11},$&\\
&&&\\
 & & $\xi^3=-a_3yz+\frac{a_2}{2} \left(z^2-y^2\right)-a_4y+c_{10}z+z\sqrt{A}\left(c_1\cos\frac{t}{2}+c_2\sin\frac{t}{2}\right) $&\\
 & & $+\sqrt{A}\left(c_5\cos\frac{t}{2}+c_6\sin\frac{t}{2}\right)-2a_2A+c_{12}$&\\
&&&\\
\hline
ND7(b)& $A=\biggl(a_3\int\sqrt{B}dx+c_1\biggr)^2$& $\xi^0=\frac{1}{\sqrt{A}} \left(c_2\cos a_3t+c_3\sin a_3t\right)+c_4,$& $a_2z-a_3y+c_5$\\
 & & $\xi^1= \frac{1}{\sqrt{B}} \left(c_2\sin a_3t-c_3\cos a_3t\right)-\frac{1}{a_3} \sqrt{\frac{A}{B}} \left(a_3y-a_2z-c_5\right)$&\\
 & & $\xi^2=a_2yz+\frac{a_3}{2} \left(z^2-y^2\right)+a_4z+c_5y+a_3\int \biggl( \sqrt{B} \int \sqrt{B} dx \biggr)dx+c_1\int \sqrt{B} dx+c_6$& \\
 & & $\xi^3=-a_3yz+\frac{a_2}{2} \left(z^2-y^2\right)-a_4y+c_5z-a_2\int \biggl( \sqrt{B} \int \sqrt{B} dx \biggr)dx-\frac{a_2c_1}{a_3}\int \sqrt{B} dx+c_7$& \\
\hline
\end{tabular}
\end{center}
\caption{\footnotesize CRCs for Non-Degenerate Ricci Tensor}
\end{table}
\end{landscape}

\begin{landscape}
\begin{table}[h]
\centering
\footnotesize
%\label{table 7}
\begin{center}
\begin{tabular}{|l|c|c|c|}
\hline
 \ \ \ \ Case  \ \ \  & Constraints  &  CRCs \ \ \  & Conformal Factor ($\lambda$)\ \ \ \  \\
\hline
\hline
ND8(a)
       &$A=k C,$  & $\xi^0=\frac{c_1}{2\sqrt{k}}(kt^2-y^2-z^2)+c_2t-\frac{c_7}{\sqrt{k}}y-\frac{c_4}{\sqrt{k}}z-a_3ty+a_2tz$&  $-\frac{C'}{2\sqrt{BC}} \biggl[\frac{c_3}{2}(kt^2+y^2+z^2)$   \\
       & where $k \in \mathbb{R}$     & $+c_3t \int \sqrt{\frac{B}{C}} dx-\frac{1}{2c_1k\sqrt{k}} \biggl(c_1 \sqrt{k}\int \sqrt{\frac{B}{C}} dx+c_{10} \biggr)^2+c_{12},$& $+(a_3y-a_2z-c_1\sqrt{k}t-c_2) \int \sqrt{\frac{B}{C}} dx$\\
       &      & $\xi^1=-\sqrt{\frac{C}{B}} \biggl[\frac{c_3}{2}(kt^2+y^2+z^2)+(a_3y-a_2z-c_1\sqrt{k}t-c_2) \int \sqrt{\frac{B}{C}} dx$& $-c_{10}t+c_8y+c_5z$\\
       &      & $-c_{10}t+c_8y+c_5z-c_3 \int \biggl(\sqrt{\frac{B}{C}} \int \sqrt{\frac{B}{C}} dx \biggr) dx -c_{11} \biggr],$& $-c_3\int \biggl(\sqrt{\frac{B}{C}} \int \sqrt{\frac{B}{C}} dx \biggr) dx-c_{11} \biggr]$\\
       &      & $\xi^2=\frac{a_3}{2}(t^2-y^2+z^2)+a_2yz+a_4z+c_1\sqrt{k}ty+c_7\sqrt{k}t+c_2y$ & $-a_3y+a_2z+c_1\sqrt{k}t$\\
       &      & $+(c_3y+c_8)\int \sqrt{\frac{B}{C}} dx+a_3 \int \biggl(\sqrt{\frac{B}{C}} \int \sqrt{\frac{B}{C}} dx \biggr) dx+c_9,$&$+c_3 \int \sqrt{\frac{B}{C}} dx+c_2$\\
       &      & $\xi^3=\frac{a_2}{2}(-kt^2-y^2+z^2)-a_3yz-a_4y+c_1\sqrt{k}tz+c_2z+c_4\sqrt{k}t$ & \\
       &      & $+(c_3z+c_5)\int \sqrt{\frac{B}{C}} dx-a_2 \int \biggl(\sqrt{\frac{B}{C}} \int \sqrt{\frac{B}{C}} dx \biggr) dx+c_6.$&\\
\hline
ND8(b)
 & $B=\frac{C^2}{4A} (\frac{A}{C})'$& $\xi^0=-\sqrt{\frac{C}{A}} \biggl[ \frac{y^2+z^2}{2} (c_8\cos t+c_9\sin t)+y(c_5\cos t+c_6\sin t)+z(c_1\cos t+c_2\sin t) $& $-\frac{C'\sqrt{\frac{A}{C}}}{C(\frac{A}{C})'} \biggl[\frac{y^2+z^2}{2} (c_8\sin t-c_9\cos t)$\\
 & & $+\frac{A}{2C} (c_8\cos t+c_9\sin t)+c_{11}\cos t+c_{12}\sin t \biggr]+c_3,$& $+y(c_5\sin t-c_6 \cos t)$\\
 & & $\xi^1=-\frac{2\sqrt{\frac{A}{C}}}{(\frac{A}{C})'} \biggl[\frac{y^2+z^2}{2} (c_8\sin t-c_9\cos t)+y(c_5\sin t-c_6\cos t)+z(c_1\sin t-c_2\cos t)$& $+z(c_1\sin t-c_2 \cos t)$\\
 & & $-\frac{A}{2C} (c_8\sin t-c_9\cos t)+c_{11}\sin t-c_{12}\cos t+\sqrt{\frac{A}{C}}(a_3y-a_2z-c_{10}) \biggr],$& $-\frac{A}{2C}(c_8\sin t-c_9\cos t)$\\
\hline
\end{tabular}
\end{center}
\caption{\footnotesize CRCs for Non-Degenerate Ricci Tensor}
\end{table}
\end{landscape}

\begin{landscape}
\begin{table}[h]
\centering
\footnotesize
%\label{table 8}
\begin{center}
\begin{tabular}{|l|c|c|c|}
\hline
 \ \ \ \ Case  \ \ \  & Constraints  &  CRCs \ \ \  & Conformal Factor ($\lambda$)\ \ \ \  \\
\hline
\hline
 & & $\xi^2=a_2yz+\frac{a_3}{2} (z^2-y^2)+a_4z+c_{10}y+a_3 \frac{A}{2C}$& $+c_{11}\sin t-c_{12}\cos t$\\
 & & $+\sqrt{\frac{A}{C}}y (c_8\sin t-c_9\cos t)+\sqrt{\frac{A}{C}} (c_5\sin t-c_6\cos t)+c_7, $&$+\sqrt{\frac{A}{C}} (a_3y-a_2z-c_{10}) \biggr]$\\
 & & $\xi^3=-a_3yz+\frac{a_2}{2} (z^2-y^2)-a_4y+c_{10}z-a_2 \frac{A}{2C}$& $a_2z-a_3y+c_{10}$\\
 & & $+\sqrt{\frac{A}{C}}z (c_8\sin t-c_9\cos t)+\sqrt{\frac{A}{C}} (c_1\sin t-c_2\cos t)+c_4, $& $+\sqrt{\frac{A}{C}} (c_8\sin t-c_9\cos t)$\\
\hline
ND8(c)
 & $A=C \biggl(-a_2\int \sqrt{\frac{B}{C}} dx+c_1 \biggr)^2$& $\xi^0=a_2 \sqrt{\frac{C}{A}}\ (c_2 \cos a_2t+c_3 \sin a_2t)+c_4,$& $-\frac{C'}{2\sqrt{BC}} \biggl[ (a_3y-a_2z-c_5) \int \sqrt{\frac{B}{C}} dx$\\
 &  & $\xi^1=- \sqrt{\frac{C}{B}} \biggl[ (a_3y-a_2z-c_5) \int \sqrt{\frac{B}{C}} dx$ & $+a_2(c_2 \sin a_2t-c_3 \cos a_2t)$\\
 & & $+a_2(c_2 \sin a_2t-c_3 \cos a_2t)-\frac{c_1a_3}{a_2}y+c_1z+\frac{c_1c_5}{a_2} \biggr],$& $-\frac{c_1a_3}{a_2}y+c_1z+\frac{c_1c_5}{a_2} \biggr]$\\
 & & $\xi^2=a_2yz+\frac{a_3}{2} (z^2-y^2)+c_5y+a_4z-\frac{c_1a_3}{a_2} \int \sqrt{\frac{B}{C}} dx$& $+a_2z-a_3y+c_5$\\
 & & $+a_3 \int \biggl( \sqrt{\frac{B}{C}} \int \sqrt{\frac{B}{C}} dx\biggr) dx+c_6,$&\\
 & & $\xi^3=-a_3yz+\frac{a_2}{2} (z^2-y^2)+c_5z-a_4y+c_1 \int \sqrt{\frac{B}{C}} dx$&\\
 & & $-a_2 \int \biggl( \sqrt{\frac{B}{C}} \int \sqrt{\frac{B}{C}} dx\biggr) dx+c_7,$&\\
\hline
\end{tabular}
\end{center}
\caption{\footnotesize CRCs for Non-Degenerate Ricci Tensor}
\end{table}
\end{landscape}

\begin{landscape}
{\Large \textbf{Appendix B}}
\begin{table}[h]
%\caption{ CRCs for Degenerate Ricci Tensor}
\centering
\footnotesize
%\label{table 9}
\begin{center}
\begin{tabular}{|l|c|c|c|}
\hline
 \ \ \ \ Case  \ \ \  & Constraints  &  CRCs \ \ \  & Conformal Factor ($\lambda$)\ \ \ \  \\
\hline
\hline
D2(a)&  $A=C$  & $\biggl( \frac{c_2}{2} \left( t^2-y^2-z^2\right)-c_4tz-c_5z-c_7ty-c_8y-c_{10}t-c_{11}\biggr) \partial_t+$  &  $\frac{A'}{2A}\ \xi^1+c_2t-c_4z-c_7y-c_{10} $  \\
 & & \ \ \ $\biggl( \frac{c_7}{2} \left( t^2-y^2+z^2\right)-c_4yz-c_{10}y+c_{12}z+c_2ty+c_8t+c_9 \biggr)\partial_y+$ &  \\
 & & $\biggl( \frac{c_4}{2} \left( t^2+y^2-z^2\right)+c_2tz+c_5t-c_{10}z-c_7yz-c_{12}y+c_{14} \biggr)\partial_z+\xi^1(x^a)\ \partial_x$&\\
 %& & $\xi^1(x^a)\ \partial_x$ & \\
  \hline
  D2(b) & $A\neq C$ and $(\frac{A'}{2A}-\frac{C'}{2C})\xi^1+h_t=0$ & $h \partial_t+c_1\left(c_3z+c_4\right) \partial_y-c_3\left(c_1y+c_2\right) \partial_z+\xi^1(x^a)\ \partial_x$ &  $\frac{A'}{2A}\ \xi^1+h_t$, where $h=h(t)$   \\
  \hline
  D3 & --- & $- \sqrt{\frac{C}{B}} \biggl[\frac{c_{10}}{2} (y^2+z^2)+c_{12}y+c_{13}z+(c_4y-c_2z) \int \sqrt{\frac{B}{C}} dx $  & $-\frac{C'}{2\sqrt{BC}}\biggl[\frac{c_{10}}{2} (y^2+z^2)+c_{12}y $  \\
& & $- c_{10} \int \left(\sqrt{\frac{B}{C}} \int \sqrt{\frac{B}{C}} dx \right) dx-c_{14} \int \sqrt{\frac{B}{C}} dx-c_{15} \biggr] \partial_x$  & $+c_{13}z+(c_4y-c_2z) \int \sqrt{\frac{B}{C}} dx$ \\
  & & $+\biggl[c_{10}y \int \sqrt{\frac{B}{C}} dx+c_4 \int \left(\sqrt{\frac{B}{C}} \int \sqrt{\frac{B}{C}} dx\right) dx +c_{12} \int \sqrt{\frac{B}{C}}dx$ & $- c_{10} \int \left(\sqrt{\frac{B}{C}} \int \sqrt{\frac{B}{C}} dx \right) dx$\\
  & & $+c_{14}y+\frac{c_4}{2} \left(z^2-y^2\right)+c_2 yz+c_5z+c_6 \biggr] \partial_y $& $-c_{14} \int \sqrt{\frac{B}{C}} dx-c_{15} \biggr]$ \\
  & & $+\biggl[c_{10}z \int \sqrt{\frac{B}{C}} dx-c_2 \int \left(\sqrt{\frac{B}{C}} \int \sqrt{\frac{B}{C}} dx\right) dx +c_{13} \int \sqrt{\frac{B}{C}}dx$ & $+c_{10} \int \sqrt{\frac{B}{C}}dx -c_4y$ \\
  & & $+c_{14}z+\frac{c_2}{2} \left(z^2-y^2\right)-c_4 yz-c_5y+c_{15} \biggr] \partial_z+\xi^0(x^a) \partial_t $& $+c_2z+c_{14}$\\
 \hline
D4 &    ---    & $h \partial_t+\xi^i(x^a)\ \partial_{x^i}; \ i=1,2,3$  &  $\frac{A'}{2A}\ \xi^1+h_t$, where $h=h(t)$  \\
  \hline
  D5 &    ---    & $h \partial_x+\xi^i(x^a)\ \partial_{x^i}; \ i=0,2,3$  &  $\frac{B'}{2B}\ h+h_x$, where $h=h(x)$  \\
  \hline
  D6 &    ---    & $c_1(c_3z+c_4) \partial_y-c_3(c_1y+c_2)\partial_z+\xi^i(x^a)\ \partial_{x^i};\  i=0,1$ & $\frac{C'}{2C}\ \xi^1$    \\
  \hline
\end{tabular}
\end{center}
\caption{\footnotesize CRCs for Degenerate Ricci Tensor}
\end{table}
\end{landscape}

\end{document}